\documentclass[a4paper]{mem}
\usepackage{natbib}
\usepackage{graphicx}
\usepackage[a4paper]{hyperref}
\idline{1000}{1}
\begin{document}
\title{Stellar populations and star formation histories in late-type dwarfs}
\author{Monica Tosi}

   \offprints{M.Tosi}
\mail{Osservatorio Astronomico, Via Ranzani 1, 40127 Bologna }

\institute{INAF - Osservatorio Astronomico di Bologna, Via Ranzani 
1, 40127 Bologna, Italy \email{tosi@bo.astro.it}}

\abstract{Studies of the resolved stellar populations in
nearby systems are crucial to understand galaxy evolution. Here, we
summarize how the interpretation of the colour-magnitude diagrams of field stars
in late-type dwarfs inside and outside the Local Group has allowed us to infer 
their star formation histories and put useful constraints on the evolution of
this type of galaxies.

\keywords{stellar populations, colour-magnitude diagrams, star formation
history, galaxy evolution} 
}

   \authorrunning{M.Tosi.}
   \titlerunning{Stellar Pops and SF histories}
   \maketitle
%

\section{Introduction}

If we aim at understanding the evolution of galaxies of whatever morphological
type, we need to follow two distinct and complementary approaches. On the one
hand we must develop theoretical models of galaxy formation, of chemical
and of dynamical evolution,  and, on the other hand, we must collect as many
and as accurate as possible observational data to constrain the models.
In particular we need to acquire reliable data on the chemical abundances,      
masses and kinematics  of the galactic components (gas, stars, dark matter), the star formation
(SF) regimes, the stellar initial mass function (IMF). It is specially important
to derive information on the actual behaviours of  SF and IMF, since they are
usually adopted as free parameters in the models. Hence, putting
stringent observational constraints on them can let us avoid unrealistic
modeling.
In galaxies close enough to let us resolve their individual stars, most of
the information on the above quantities can be derived by observing their resolved
stars. To infer the SF regime and history of resolved stellar populations, the 
best tool is their colour-magnitude diagram (CMD), because it shows the         
signature of their evolutionary status.

   \begin{figure*}
   \centering
   \includegraphics[width=13cm,clip]{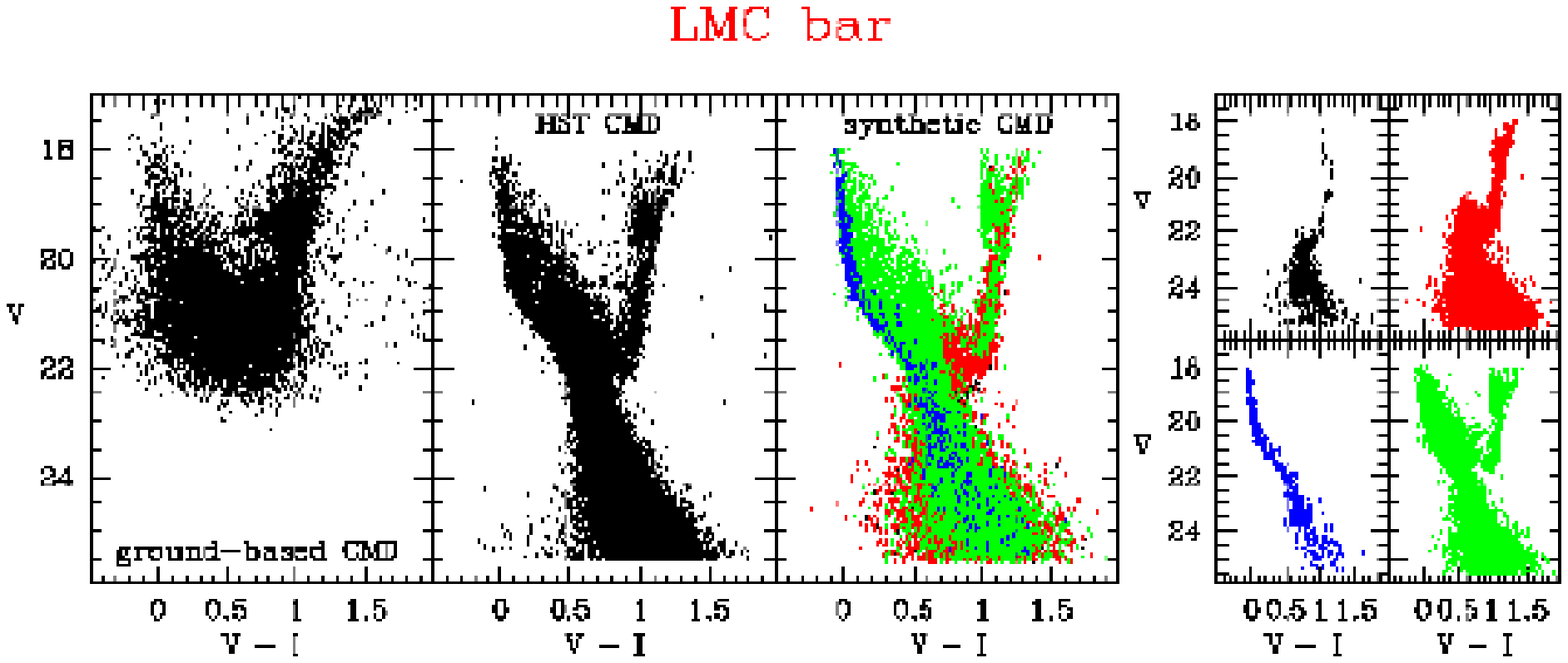}
     \caption{CMDs of one field of the LMC bar. Left-hand panel: ground based
     CMD from the 1.5 m Danish telescope \citep{Clem03}. Second panel from
     left: CMD of the same field as derived by \citet{Sme02} from
     HST/WFPC2 photometry; third panel from left: from the Coimbra 
     experiment, synthetic CMD by \citet{T02}. Right-hand panel,
     the synthetic CMD of the previous panel splitted in its four
     episode components, from the oldest to the youngest one clock-wise from the
     top-left panel. }  
        \label{lmc}
    \end{figure*}

\section {The synthetic CMD method}

To derive the SF history of dwarf galaxies, several years ago 
Laura Greggio and myself developed a method
\citep{T91,G98} based on the         
comparison of empirical and synthetic CMDs.
The synthetic CMDs are created via MonteCarlo extractions on homogeneous sets of
stellar evolution tracks. They take into account all the theoretical parameters
(IMF, age, metallicity, small number statistics, etc.), must contain the same
number of stars as the observational CMD (or portions of it), 
and must be affected by the same
photometric error, incompleteness and blending factors. Hence, a combination
of theoretical parameters is acceptable only if the resulting CMD reproduces all
the features of the observational one: morphology, colours, luminosity functions
(LF), number of stars in specific evolutionary phases.
The comparison between empirical and synthetic CMD allows us to evaluate   
whether or not the parameter combination of the latter is acceptable. By
checking all the combinations we can derive the epoch, duration, intensity of
the SF episodes, number of episodes and quiescent intervals, IMF, metallicity.

Since this is a statistical approach, we cannot pretend to get unique solutions
for the SF history of the examined region, but we can strongly reduce
the range of possible scenarios.

   \begin{figure*}
   \centering
   \includegraphics[width=13cm,height=9cm,clip]{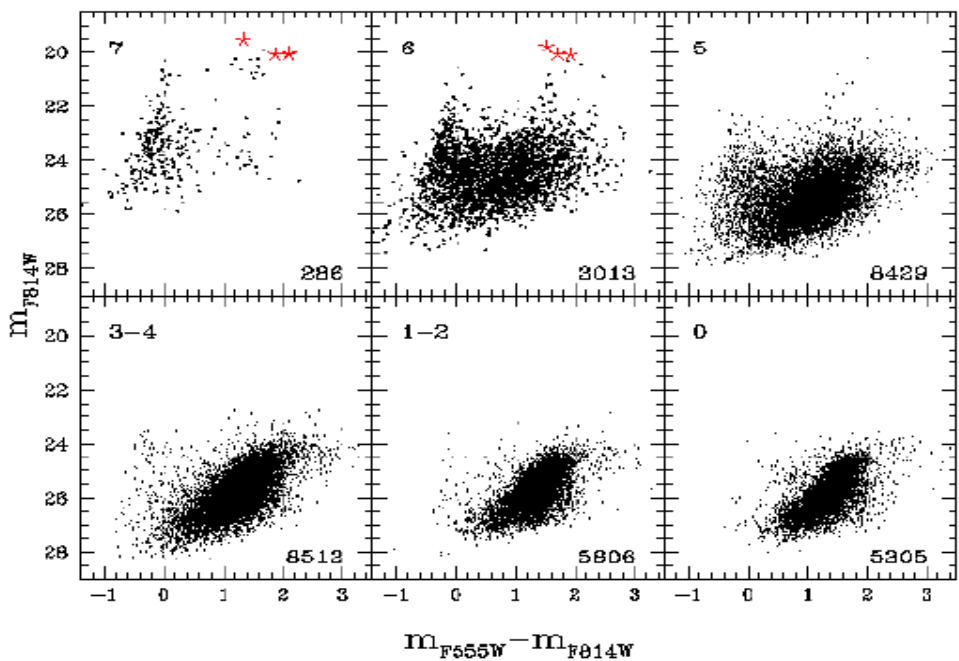}
     \caption{CMDs of concentric regions of the BCD galaxy NGC 1705,
     as derived from HST/WFPC2 photometry \citep{T01}. The number in the  
     top-left corner of each panel indicates the region, 
     from the central one (7) to the outermost (0). The number of resolved stars
     in each region is shown in the bottom-right corner.
      Asterisks represent possible star clusters. }  
        \label{cmd1705}
    \end{figure*}

   \begin{figure*}
   \centering
   \includegraphics[width=13cm,height=8cm,clip]{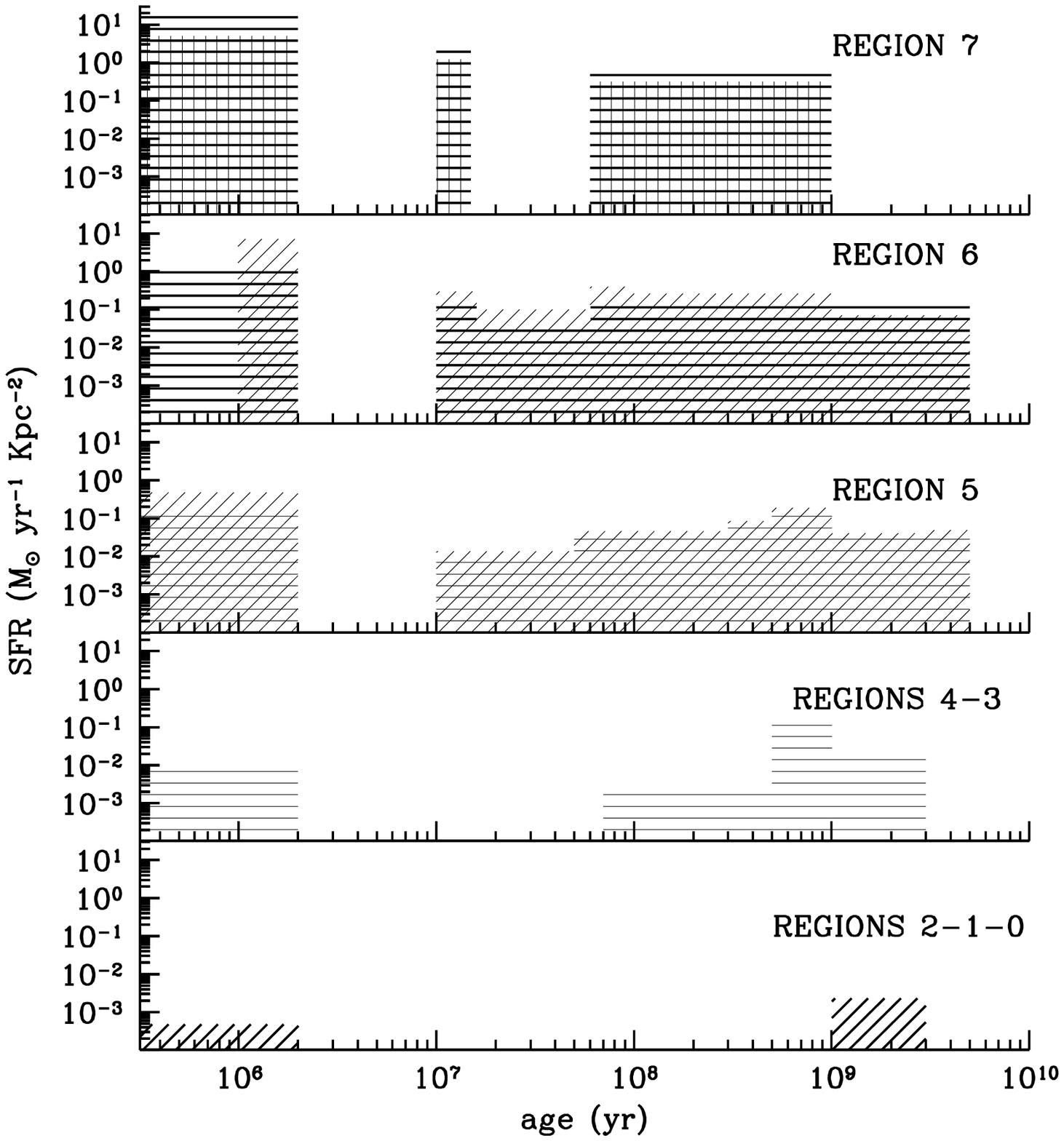}
     \caption{SF rate per unit area as a function of age of the various regions
      of NGC1705, as derived from the
     application of the synthetic CMD method to the diagrams of
     Fig.\ref{cmd1705} by \citet{annibali}. Notice that both axes are on
     logarithmic scale. The different inclination angles of the histogram
     filling show the SF history derived assuming different IMF slopes:
     Horizontal lines refer to Salpeter's IMF (i.e. exponent $\alpha$=2.35),
     vertical lines to $\alpha$=2.2, and slanted lines to $\alpha$=2.6.} 
    \label{sf1705}
    \end{figure*}

When we started working in this field, the CMDs were obtained from
relatively small, relatively old telescopes, such as the 1.5m Danish and the    
2.2m ESO/MPI in La Silla \citep{F89,T91} or the 2.5m INT in the 
Canary Islands \citep{Ga96}.
 When the first CMDs resulting from HST photometry became
available, they appeared impressively deeper and tighter than the previous 
ones (see e.g. Fig.\ref{lmc}) and triggered so much interest on the synthetic 
CMD method, that many
new groups developed their own procedures to apply it [e.g. \cite{TS96}].
To check if the scenarios resulting from the approaches of different groups are
consistent with each other, Carme Gallart set up a very interesting experiment
which was held in 2000 \citep{SG02}  in Coimbra 
(Portugal). Each participating group was given the empirical
CMD of a field on the bar of the LMC as derived by \citet{Sme02} 
from HST-WFPC2 photometry (second panel from left in Fig.\ref{lmc})
and the catalogue of the artificial star
test kindly performed by A.Dolphin to get photometric errors, incompleteness and
blending factors. A referee (E.Skillman) was in charge of
comparing the results of the different groups and of summarizing them all. 
About ten groups participated \citep{GAB, Rizzi, Cole, Dolphin, DVG, Holtzman,
Harris, T02}. The experiment was very successful, because the various scenarios
turned out to be indeed consistent with each other, thus giving support to the
method reliability, and because we had the
opportunity to compare the different procedures with each other and evidentiate
strengths and weaknesses of each one. The SF regime in that LMC field turned
out to have been fairly continuous over the whole Hubble time, with no
apparent interruptions (at least at more recent epochs when time
resolution is higher), but with significant ups and downs in the SF rate (see
the filled histogram in the right-bottom panel in Fig.\ref{sfall}). This SF
history is significantly different from that presented by \citet{PT98}
for LMC clusters, represented in Fig.\ref{sfall} by the empty histogram. This
difference reveals how important it is to derive the SF history of both cluster
and field stellar populations in as many galactic regions as possible.

\section {The SF histories of late-type dwarfs}

Nowadays the SF history has been inferred in more or less detail for all the
late-type dwarfs in the Local Group. As reviewed e.g. by Grebel (1998), all
these galaxies show what we \citep{T91} defined as a {\it gasping} SF
regime (i.e. long episodes of moderate SF activity, possibly separated by
short quiescent phases), rather than a bursting one (short and intense
episodes of SF activity, separated by long quiescent phases).

Local Group galaxies are obviously the best ones to accurately derive the
SF history back to the oldest epochs. In more distant galaxies, crowding
and magnitude limit make the fainter/older stars increasingly difficult to
resolve and, correspondingly, the lookback time reachable by the photometry
is increasingly short (ranging between a few 10$^9$ yr to a few 10$^8$ yr).
None the less, we have to study galaxies also outside the Local Group, because we
do know that not all the morhological types are present in the Group. Indeed,           
ellipticals and blue compact dwarfs (BCDs), i.e. the most and the least evolved 
galaxies can be found only outside it.

Our group is studying a number of external late-type dwarfs of particular 
interest: NGC1569, a dwarf irregular at 2.2 Mpc, with 
particularly strong SF 
activity, three super-star-clusters, and evidence of galactic winds
\citep{DM97,G98,aloisi01,O01},
NGC1705, a BCD at 5.1 Mpc, with one super-star-cluster and evidence of 
galactic winds \citep{T01,annibali},
IZw18, a BCD at 10-14 Mpc which is the most metal-poor galaxy discovered so
far \citep{aloisi99}, and a few more, still to be observed with the HST/ACS.

\begin{table*}
\caption{BCDs already studied with the synthetic CMD method}
\begin{tabular}{l|c|c|c|l}
\hline

Galaxy & D (Mpc) & SFR (M$_{\odot}$ yr$^{-1}$) & 12+log(O/H) & Reference\\
\hline
 I~Zw~18    & 10-12 & 3-10~10$^{-3}$     & 7.18 & \citet{aloisi99}\\
 VII~Zw~403 & 4.4   & 1.3~10$^{-2}$      & 7.69 & \citet{L98} \\
 UGCA~290   & 6.7   & 1.1~10$^{-2}$      &  ?   & \citet{c00}, (2002)\\
 I~Zw~36    & 5.8   & 2.5~10$^{-2}$      & 7.77 & \citet{sl01}\\
 NGC~6789   & 3.6   & 4.0~10$^{-2}$      & 7.7? & \citet{dr01} \\
 UGC~5272   & 5.5   & 6~10$^{-3}$        & 7.83 & Hopp et al., in prep\\
 MrK~178    & 4.2   & $\le$~~10$^{-2}$   & 7.95 & \citet{sl00}\\
 NGC~4214   & 2.7   & 8~10$^{-2}$        & 8.27 & \citet{dr02}\\
 NGC~1569   & 2.2   & 5~10$^{-1}$        & 8.31 & \citet{G98}\\
 NGC~1705   & 5.1   & 1~10$^{-1}$        & 8.36 & \citet{annibali} \\

\hline
\end{tabular}
\end{table*}

The case of NGC1705 is particularly instructive, because the photometry was
deep and good enough to let us resolve its stars from the most central regions
to the extreme outskirts. We have thus been able to divide the galaxy in 8
roughly concentric regions, all sufficiently populated by individual stars,
and derive the SF history of each region. Fig.\ref{cmd1705}  shows the CMDs of 
the various regions: it is apparent that young massive stars are concentrated
at the center and their percentage rapidly decreases outwards, while faint
red stars are increasingly visible towards the outer regions. The latter
circumstance doesn't necessarily imply that old stars are absent at the center,
it simply means that crowding is too severe there to let us resolve them.
In the outer regions, where crowding is definitely not a problem, the CMDs
present a well defined upper portion of the red-giant branch (RGB), whose
tip is also very well defined and allowed us to accurately derive the galaxy 
distance \citep{T01}.

By applying to each region the synthetic CMD method, we \citep{annibali}
have inferred their SF histories, summarized in Fig.\ref{sf1705}. There. the SF
rate per unit area is plotted as a function of age. It can be seen that, except
for the innermost region, where crowding does not allow us to reach old lookback
times, all the regions have been forming stars since at least 5 Gyr.
On average, the SF appears to have been rather continous:
There are evidences for interruptions in the SF activity, but always shorter
than a few Myr or tens of Myr, at least in the age range where we do have this
time resolution (i.e. in the last 1 Gyr or so). Quiescent phases of 100 Myr or
longer would have appeared as gaps in the empirical CMDs of stars younger than
1 Gyr, and such gaps are absent. The SF history of NGC1705 shows three striking
features: one is the burst occurred in the central regions 10--15 Myr ago, when
the SSC also formed and when the observed galactic wind is supposed to have
originated; the second is the quiescent phase with no SF anywhere in the galaxy
right after such burst, probably due to the hot gas shocks and winds caused by
the explosions of the burst supernovae; and the third is
the new, even stronger, SF activity occurring everywhere in NGC1705 (but much
higher in the inner regions) in the last 2 Myr. This latter event puts
interesting constraints on the cooling timescales of the gas heated by the
supernovae generated in the previous burst and on the modeling of SF processes.

   \begin{figure*}
   \centering
   \includegraphics[width=13cm,clip]{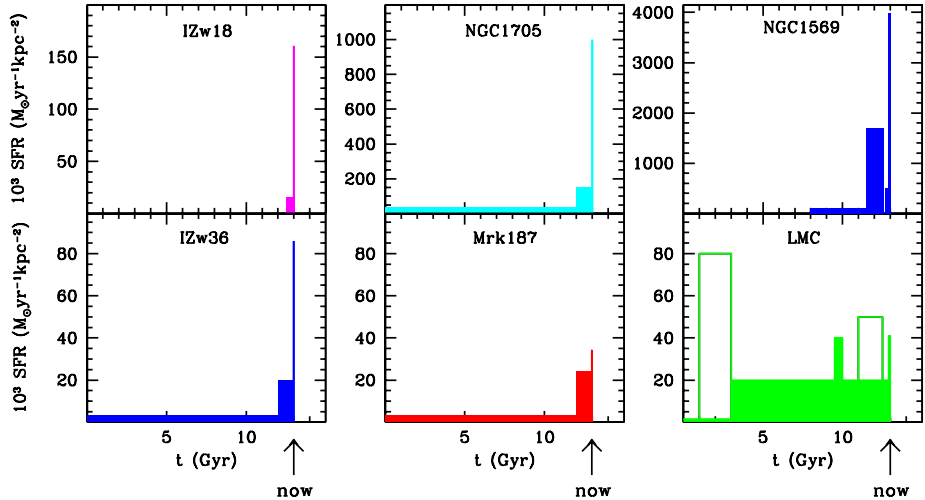}
     \caption{SF rate per unit area as a function of time for some of the
     late-type dwarfs studied so far with the synthetic CMD method (see Table 1
     for references). Notice
     that IZw18, IZw36, Mrk187 and NGC1705 are classified as BCDs, while
     NGC1569 is classified as a dwarf irregular and the LMC as a giant
     irregular. 
     }  
        \label{sfall}
    \end{figure*}

SF histories qualitatively similar to that described here for NGC1705 have been
derived by us also for NGC1569 and for IZw18 and by other people for other
BCDs, also observed with HST and studied with the synthetic CMD method. Table 1
lists all the BCDs outside the Local Group whose SF history has been
inferred with this method. Fig.\ref{sfall} shows the resulting diagrams of the
SF rate per unit area vs time for some of the BCDs and for two representative
irregulars: NGC1569 and the LMC. It is interesting to note that the galaxy with
stronger SF activity is the dwarf irregular, NGC1569, the only one with a rate
comparable to the 1 $M_{\odot}yr^{-1}$ required by \citet{BF96}
models to let a late-type galaxy contribute to the blue galaxy excess observed
in counts at intermediate redshift. All the other dwarfs, independently of being
classified as BCDs or irregulars present much lower SF rates. The only apparent
difference in the SF of the two type of dwarfs is the presence of a very recent
burst in BCDs. This is presumably due to  selection effects that made it
impossible to discover dwarfs without recent bursts beyond a certain distance.

From all the studies on the SF histories of late-type dwarfs performed so far,
we can schematically summarize the main results:
\begin{itemize}
\item no evidence of long interruptions in the SF activity has been found in any
late-type dwarf;
\item strong bursts don't seem to be frequent (NGC1569 showing the highest SF
rate, followed by NGC1705 with a rate a factor of two lower);
\item no galaxy currently experiencing its first SF activity has been found yet
(all the studied ones were already active at the lookback time reached by the
photometry);
\item the SF regime seems to be a gasping rather than a bursting one in all   
kinds of late-type dwarfs, both in the Local Group and outside it;
\item no significant difference has been found in the SF histories and in the
stellar populations of BCDs and irregulars, except that the former ones have a
recent SF burst.
\end{itemize}

These results suggest, in my opinion, that late-type dwarfs are unlikely
significant contributors to the excess of blue galaxy counts at red-shift around
0.7 -- 1, and that, in spite of being poorly evolved, they should not be
considered young galaxies.

\begin{acknowledgements}
  I'm grateful to all the persons who are, or have been, involved in this long
  term project and, in particular, to Laura Greggio, Alessandra Aloisi and 
  Francesca Annibali. Regina Schulte-Ladbeck and Ulrich Hopp have kindly 
  provided the data for Table 1 and Fig.\ref{sfall}.
  This work has been partially supported by the Italian ASI and MIUR through    
  grants  ARS-99-44, ASI-I/R/35/00 and Cofins 2000 and 2002.
\end{acknowledgements}

\bibliographystyle{aa}

\end{document}